\documentclass[12pt]{article}
\usepackage[T1]{fontenc}
\usepackage[utf8]{inputenc}
\usepackage[english]{babel}
\usepackage{orcidlink}
\usepackage{bm}
\usepackage{amsmath}
\usepackage{color}
\usepackage{graphicx}
\usepackage{caption}
\usepackage{subfig}
\usepackage{amssymb}
\usepackage{float}
\usepackage{cite}
\usepackage{hyperref}
\usepackage[makeroom]{cancel}
\usepackage{booktabs}
\usepackage{multirow}

\graphicspath{{./plots/}} 

\definecolor{blue}{rgb}{0.27, 0.42, 0.81}
\usepackage{hyperref}
\hypersetup{
colorlinks, linktocpage, bookmarksnumbered,
pdfstartview=FitH,
citecolor=blue,
urlcolor=blue,
linkcolor=red,
filecolor=blue
}

\begin{document}
\allowdisplaybreaks

\begin{titlepage}

\begin{flushright}
{\small
USC-TH-2024-04\\
\today \\
}
\end{flushright}

\vskip1cm
\begin{center}
{\Large \bf\boldmath Finite width effects in nonleptonic D-decays}
\end{center}

\vspace{0.5cm}
\begin{center}
{Girish Kumar \orcidlink{0000-0001-6051-2495},
Sudheer Muhammad \orcidlink{0009-0005-1136-0057}, 
Alexey A. Petrov \orcidlink{0000-0002-4945-4463}} \\[7mm]
{\emph{Department of Physics and Astronomy, \\
University of South Carolina, Columbia, \\
South Carolina 29208, USA}}\\[0.3cm]
\end{center}

\vspace{7mm}
\begin{abstract}
\noindent
Many analyses of two-body non-leptonic decays of $D$-mesons rely on flavor SU(3) symmetry relations and fits of experimental data of decays rates to extract the universal transition amplitudes. Such fits assume that the final state mesons are well-defined asymptotic states of QCD. We develop a technique to take into account the finite width effects of the final state mesons and study their effects on the extracted values of transition amplitudes.
\end{abstract}
\end{titlepage}


\section{Introduction}
Charm transitions play an important role in flavor physics. The availability of large statistical samples of charm data implies that studies of the decays of charmed mesons could yield interesting constraints on New Physics models or serve as an important laboratory for our understanding of non-perturbative QCD.  

As charm quark's mass is ``in-between'' being heavy and light, as compared to $\Lambda_{\rm QCD}$, the techniques developed for dealing with non-perturbative QCD effects in heavy or massless quark limits are not directly applicable to non-leptonic decays of the charmed states. Some limited success in computing decay amplitudes and CP-violating asymmetries is enjoyed by the QCD Sum Rule techniques, although such an approach requires additional assumptions. Finally, lattice QCD has been applied to computing non-leptonic decays of charmed mesons. Although the computations are still quite expensive in terms of computing time at the moment, there is steady progress. With new experimental data expected from the BESIII, LHCb, and Belle II experiments, it is essential to understand the transition amplitudes describing decays and mixing of charmed mesons.

In the absence of model-independent techniques, a popular approach is to use flavor symmetries to relate the experimental data on two-body decays to a few universal parameters, which could then be used to predict the yet-to-be measured decay rates \cite{Petrov:2021idw}. The approach is analogous to choosing a basis in the space of SU(3) amplitudes. There are two main approaches to choosing such a basis. One employs the Wigner-Eckart theorem to write each decay amplitude in terms of the product of the universal reduced matrix elements and process-dependent Clebsch-Gordan coefficients. The second approach employs a set of physics-motivated flavor-flow amplitudes that serve as process-independent base vectors. Both approaches were shown to be equivalent both in the flavor SU(3) limit and in taking into account the SU(3) breaking effects \cite{Muller:2015lua}. 

All fits of flavor $SU(3)_F$ amplitudes in charmed particle decays assume that the final-state mesons or baryons are proper asymptotic states. This is, however, not a very good approximation for the light-quark vector and scalar mesons produced in $D$ decays whose widths could be rather large and significantly different from one another. Those widths effectively introduce another source of $SU(3)$-breaking, which must be considered. The possible experimental effect could be rather substantial, especially for the pseudoscalar-vector and vector-vector final states, as the effect scales as $\Gamma_f/E_{\rm rel}$, where $\Gamma_f$ is the width of a final state particle, while $E_{\rm rel}=m_D-\sum_f m_{f}$ with $m_f$ is the mass of a final state particle. Interestingly, some of the $D$ decays to $VV$ final states are only allowed because of the finite widths of the final state particles. So we expects a rather substantial effect on the numerical values of extracted amplitudes, which implies that such effects must be taken into account. 

In what follows, we will consider the effects of the finite widths of the final-state mesons on extracting the decay amplitudes. As the goal of this paper is to demonstrate that the finite widths of the final state particles can affect the values of the extracted universal transition amplitudes, there is no difference in the formalism chosen to parameterize the decay amplitudes. We will choose the flavor flow diagram approach and neglect the SU(3) breaking effects in the amplitudes. We will show that the finite widths effects affect the extracted values of the decay amplitudes, although the effect is rather small in $D \to PP$ and $D \to PV$ transitions. We will consider the fits in Section \ref{AmplitudeFits} and conclude in Section \ref{Conclusions}.

\section{Decay amplitudes and phase space}
\label{AmplitudeFits}
The problem of consistently considering the finite widths of particles in the computations of production cross-sections has been approached with various tools \cite{Beneke:2015vfa}. Approaches such as effective theories \cite{Beneke:2003xh}, complex mass scheme (CMS) \cite{Denner:2014zga}, etc., have been in use for the processes involving the gauge $W$ and $Z$ bosons and the top quark. Several important issues dealing with unstable states, such as unitarity, have been previously addressed \cite{Veltman:1963th}. 

The problem of considering the finite particle widths in non-leptonic D-decays is more straightforward, as one does not need to worry about such issues as gauge invariance. Moreover, as the decay amplitudes are not computed but fitted to the experimental data, one must consider modification of the final state phase space by the final state particles' widths. The effect is important not only in the fits of the decay amplitudes, but also in computing the mass and width differences in the exclusive approaches to $D^0-\overline{D^0}$ mixing parameters \cite{Falk:2001hx}.

\begin{figure}[t]
\centering
     \includegraphics[width=6.8cm]{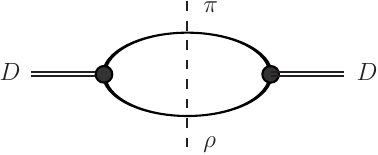}
\caption{An example of the imaginary part of the amplitude squared for two-body D-decays. An amplitude to the right of the cut, denoted by the vertical dashed line, must be taken as hermitian-conjugate of the amplitude to the left of the cut.}
\label{Fig1}
\end{figure}

A convenient way to compute a two-body decay width $\Gamma(D \to f)$, at least in the approximation of $\Gamma_f=0$, is to use unitarity in the form of Cutkosky rules \cite{Cutkosky:1960sp}. In this approach, computing the imaginary part of the forward scattering amplitude ${\cal M}(D \to D)$ gives the partial width $\Gamma(D \to f)$ of the transition, as seen from Fig.~\ref{Fig1}. We will modify this formalism to include the non-zero widths of the final state particles. To do so, for each unstable particle, we will introduce a width by introducing complex masses, defined as the locations of the poles in the complex $p^2$ plane of the corresponding propagators with momentum $p$, i.e., by shifting the pole associated with the final state particles $m_f^2 \to m_f^2 - i \Gamma_f m_f$. This scheme is similar to the complex mass scheme.

It is important to point out that since unstable particles are not asymptotic particle states, their propagators should not be cut in the application of the Cutkosky formalism. In practice, however, at least in the narrow-width approximation, this fact is often ignored, so unstable particles are often treated as if they are stable. Following Ref. \cite{Donoghue:2019fcb}, we show (see Appendix \ref{Appendix}) that, at least to the precision we work here, it is appropriate to explicitly take into account the non-zero width of the cut light final state meson line leads to corrections in the extracted values of decay amplitudes. 

{The optical theorem} relates the imaginary part of the forward scattering amplitude, ${\cal M}_\text{loop}(D\to D)$, associated with the loop diagram to the rate of particle $D$ decaying to final state $f$ as
\begin{align}\label{eq:optical-th}
	\Gamma(D \to f) = \frac{1}{m_D}\operatorname{Im}{\cal M}_\text{loop}(D\to D).
\end{align}
In this article, we will focus on the Cabibbo-favored (CF) two-body decays of $D$ mesons,
where final state contains either a pair of pseudoscalars ($f= P_1 P_2$)
or a pseudoscalar and a vector meson ($f=P V$).
For the former, we define the $D$-$P_1$-$P_2$ vertex factor as $i {\cal A}_{D\to P_1P_2}$,
to calculate the imaginary part of the loop diagram in Fig.~\ref{Fig1} (with $P_1$ and $P_2$ in the loop).
Using Eq.~\eqref{eq:optical-th}, we obtain:
\begin{align}\label{eq:decay-rate-D2PP}
	\Gamma(D \to P_1P_2)
	= -\frac{|{\cal A}_{D\to P_1P_2}|^2}{16\pi^2\, m_D}
	\operatorname{Im}\int_0^1 dx \log \Delta (x),
\end{align}
with $\Delta (x) = \tilde m_{P_1}^2-x\{\tilde m_{P_1}^2-\tilde m_{P_2}^2+m_{D}^2(1-x)\}$, where we use the notation  
 $\tilde m_{a}^2 \equiv  m_a^2 - i m_a \Gamma_a$ underlining the shift in the pole location due to the
inclusion of finite width of final state particles as discussed earlier. 

In the limit of asymptotic states, i.e., $\Gamma_{P_1}$, $\Gamma_{P_2} \to 0$, the integral in
Eq.~\eqref{eq:decay-rate-D2PP} is complex for $\Delta(x)<0$ only.
Then, using the identify $\log (-x) = \log |x| - i \pi $, Eq.~\eqref{eq:decay-rate-D2PP} simplifies to
\begin{align}\label{eq:decay-rate-D2PP-0}
	\Gamma(D \to P_1P_2)
	&= \frac{|{\cal A}_{D\to P_1P_2}|^2}{16\pi\, m_D}
	\int_0^1 dx \,{\theta}\left(x\{m_{P_1}^2-m_{P_2}^2+m_{D}^2(1-x)\}- m_{P_1}^2\right),\nonumber\\
	&= \frac{p^\ast }{8\pi\, m_D^2} |{\cal A}_{D\to P_1P_2}|^2 \,\theta(m_D - m_{P_1}-m_{P_2}),
\end{align}
recovering the typical decay rate formula for $D \to P_1 P_2$ processes used in literature \cite{Bhattacharya:2009ps}.
Here, $p^\ast$ is the magnitude of the three-momentum of each final state particles in the
rest frame of $D$-meson, given as
\begin{align}\label{eq:3-momentum}
	p^\ast = \frac{\sqrt{\lambda (m_D^2, m_{P_1}^2, m_{P_2}^2)}}{2\, m_D},
\end{align}
where $\lambda(x, y, z) = x^2 + y^2 +z^2 -2(xy + yz + zx)$ denotes the K\"all\'en function.

The case of $D\to PV$ decays is more involved as polarization of the vector meson has to be taken into account \cite{Bhattacharya:2008ke,Cheng:2010ry,Cheng:2024hdo}. Defining the four-momentum of $D$, $P$, and $V$ mesons as $p_D$, $p_P$, and $p_V$,
and the polarization vector of $V$ meson as $\epsilon(p_V)$, one can write the most general Lorentz decomposition of the decay amplitude for $D\to PV$ as:
\begin{equation}\label{eq:D2PV-Lorentz}
	\begin{aligned}
	{\cal M}[D(p_D) \to P(p_P) V(p_V)]
	&= (A\, p_D^\mu + B\, p_P^\mu) \epsilon_\mu(p_V) {\cal A}_{D\to PV},\\
	&= (A+B)\, p_D^\mu   \epsilon_\mu(p_V) {\cal A}_{D\to PV},
\end{aligned}
\end{equation}
where coefficients $A, B$ are Lorentz scalar quantities. To arrive at the the second line, we have used the relation $p_D = p_P + p_V$ and the Ward identity $\epsilon \cdot p_V = 0$. From Eq.~\eqref{eq:D2PV-Lorentz}, we find  the $D$-$P$-$V$ vertex factor to be $i (A+B)\, p_D^\mu  {\cal A}_{D\to PV}$.

Now, similar to the $D\to P_1P_2$ case, we can include the width effects of final state particles in the
phase space by calculating the imaginary part of the loop diagram shown in Fig.~\ref{Fig1}.
Taking the vector meson propagator in the complex mass-scheme
\begin{equation}
	i\frac{-g^{\mu\nu} + \frac{k^\mu k^\nu}{m_V^2 - im_V \Gamma_V}}{k^2 - m_V^2 + im_V \Gamma_V},
\end{equation}
and using Eq.~\eqref{eq:optical-th}, we obtain the following decay rate formula for $D\to PV$:
\begin{align}\label{eq:decay-rate-D2PV}
	\Gamma(D \to PV)
	&= \frac{(A+B)^2\, m_D}{4\pi^2 m_V^2}|{\cal A}_{D\to PV}|^2\nonumber \\
	&\hskip1cm \times \operatorname{Im}\int_0^1 dx\left[\frac{\Delta(x)}{2}
	+ \left(m_V^2-m_D^2 x^2 - \frac{\Delta(x)}{2}\right)\log\Delta(x)\right],
\end{align}
where the function $\Delta(x)$ is similar to the one defined below Eq.~\eqref{eq:decay-rate-D2PP}\
after the substitutions $\{P_1, P_2\} \to \{V, P\}$.
In the $\Gamma_V$, $\Gamma_P \to 0$ limit, again the logarithmic term 
contributes to the imaginary part only when $\Delta(x)< 0$, yielding   
\begin{align}\label{eq:decay-rate-D2PV-0}
	\Gamma(D \to PV)
	&= \frac{(A+B)^2\, (p^{\ast})^3}{8\pi m_V^2}|{\cal A}_{D\to PV}|^2,
\end{align}
where $p^{\ast}$ is the magnitude of the three-momentum of $P$ and $V$ and given by Eq.~\ref{eq:3-momentum}
after obvious label substitutions. With the choice $A+B = m_V/m_D$, one obtains the well-recognized
expression of $D\to PV$ decay rate used in literature for asymptotic final states \cite{Bhattacharya:2008ke}.

We are now ready to perform fits to flavor $SU(3)_F$ amplitudes in charm decays, taking into account finite width effects of final state particles using the decay rate formulas in Eqs.~\eqref{eq:decay-rate-D2PP-0} and
\eqref{eq:decay-rate-D2PV}.

\subsection{D-decays to two pseudoscalars}
\begin{table}[t]
\centering
\caption{Amplitude decomposition of CF
decays of charmed mesons to a pair of pseudoscalars in terms of
$SU(3)_F$ flavor-topological diagrams (see, for example, Ref.~\cite{Bhattacharya:2009ps})
together with corresponding measured branching ratios (taken from
Ref.~\cite{ParticleDataGroup:2024cfk}) and theoretical predictions
based on best-fit solution given in Table~\ref{tab:D2PP-best}. 
\label{tab:D2PP-CF}}
\begin{tabular}{c c c c c}
\toprule
Meson & Mode & Amplitude & ${\cal B}_\text{exp}$ ($\%$) & ${\cal B}_\text{fit}$ ($\%$) \\
 \midrule
$D^0$ &$K^- \pi^+$  &  $T+E$ &$3.947\pm0.030$ & $3.947 \pm 0.075$\\
       &$\overline{K}^0 \pi^0$    &  $\frac{1}{\sqrt{2}}(C-E)$
 			&$2.311\pm0.036$ & $2.312 \pm 0.035$\\
       &$\overline{K}^0 \eta$   &  $\frac{1}{\sqrt{3}}C$
			 &$0.958\pm0.002$ & $0.958 \pm 0.003$\\
       &$\overline{K}^0 \eta^\prime$& $-\frac{1}{\sqrt{6}}(C+3E)$ &$1.773\pm0.047$ & $1.772 \pm 0.043$\\
 \midrule
$D^+$ &$\overline{K}^0 \pi^+$  & $C+T$ & $3.067\pm 0.053$ & $3.067 \pm 0.086$\\
 \midrule
$D^+_s$&$\overline{K}^0 K^+$ &$C+A$ &$2.202\pm0.060$ &$2.204 \pm 0.106$\\
 		&$\pi^+ \eta$ & $\frac{1}{\sqrt{3}}(T-2A)$&$1.68\pm0.09$ & $1.68 \pm 0.09$\\
 		&$\pi^+ \eta^\prime$  &$\frac{2}{\sqrt{6}}(T+A)$ &$3.94\pm0.25$ & $3.97 \pm 0.12$\\
\bottomrule
\end{tabular}
\end{table}

We consider eight CF decay modes of charm mesons to two pseudoscalars, listed in Table~\ref{tab:D2PP-CF}. Their amplitudes in the topological flavor-flow approach can be decomposed in terms of four tree amplitudes: color-favored ($T$), color-suppressed ($C$), exchange ($E$), and annihilation ($A$) amplitude, as shown in Table~\ref{tab:D2PP-CF}.
Note that regarding decays involving $\eta$, $\eta^\prime$ states, we follow
$\eta-\eta^\prime$ mixing,
\begin{align}\label{eq:eta-etap}
\begin{pmatrix}
  \eta  \\
  \eta^\prime
\end{pmatrix} =  -\begin{pmatrix}
  \cos \theta & \sin \theta   \\
   \sin \theta & -\cos \theta 
   \end{pmatrix}\begin{pmatrix}
  \eta_8  \\
  \eta_1
\end{pmatrix},
	\end{align}
where $\eta_1$ and $\eta_8$ are the $SU(3)$ singlet and octet states, respectively, defined as
\begin{align}
	\eta_1 &= \frac{1}{\sqrt{3}}\left(u\bar u + d \bar d +  s \bar s\right),\\
	\eta_8 &= \frac{1}{\sqrt{6}}\left(u\bar u + d \bar d - 2 s \bar s\right).
\end{align}
In our analysis we take $\theta = \arcsin(1/3)$ \cite{Bhattacharya:2021ndt} (for use of other prescriptions for $\eta-\eta^\prime$ mixing, see \cite{Bhattacharya:2010uy,Bolognani:2024zno}).

The topological amplitudes  contain complex phases, but since one overall phase is spurious, we will take $T$ to be real. Therefore, there are total $7$ parameters to determine from fit: the magnitude of amplitudes\footnote{For brevity, we use the same notation to denote amplitude magnitudes.} ($T, C, E, A$) and phases ($\phi_C, \phi_E, \phi_A$). To this end we perform a $\chi^2$ minimization defined as
\begin{align}\label{eq:chi2}
	\chi^2 = \sum_{i=1}^8 \frac{({\cal B}_i^\text{th} - {\cal B}_i^\text{exp})^2}
						{(\Delta {\cal B}_i^\text{exp})^2},
\end{align}
where ${\cal B}_i^\text{th}$ and ${\cal B}_i^\text{exp}$ denote the theoretical and experimental mean of branching ratios, respectively, and $\Delta {\cal B}_i^\text{exp}$ the corresponding  $1\sigma$ experimental uncertainty. The experimental data is provided in Table~\ref{tab:D2PP-CF},  while the theoretical branching ratios, which depend on $7$ topological amplitude parameters, are evaluated using Eq.~\ref{eq:decay-rate-D2PP}. We use python package \texttt{IMINUIT}\cite{iminuit,James:1975dr} to minimize the $\chi^2$ function and find the best-fit solution. The obtained solution yields a vanishing $\chi^2_\text{min}$ value and is given in Table~\ref{tab:D2PP-best}. From the fit, we note that $T$ is the largest amplitude while $A$ the smallest. We do not find any appreciable change in fit results compared to those without including width effects of final state pseudoscalars. This is expected as widths of these particles \cite{ParticleDataGroup:2024cfk} are negligibly small.

The predicted branching ratios based on the solution in Table~\ref{tab:D2PP-best}
are listed in the last column of Table~\ref{tab:D2PP-CF},
indicating the excellent agreement between theory and $D\to P_1P_2$ data.  
\begin{table}[t]
\centering
\caption{Fit results for $D\to PP $ decays (the amplitudes are
in units of $10^{-6}$ GeV).
\label{tab:D2PP-best}}
\begin{tabular}{c c}
\toprule
\textbf{Parameters} & \textbf{Solution} \\
\midrule
$T$ & $2.892 \pm 0.020$ \\
$C$ & $2.285 \pm 0.002$ \\
$\phi_{C}$ & $(-151.321 \pm 0.371) ^\circ$ \\
$E$ & $1.531 \pm 0.021$ \\
$\phi_{E}$ & $(118.915 \pm 0.615) ^\circ$ \\
$A$ & $0.653 \pm 0.056$ \\
$\phi_{A}$ & $(67.230 \pm 2.980) ^\circ$ \\
\midrule
$\chi^2_\text{min}$ & $0.018$ \\
\bottomrule
\end{tabular}
\end{table}
%

\subsection{D-decays to a pseudoscalar and a vector}
\begin{table}[t]
\caption{Diagrammatic description of CF
decays of charmed mesons to one pseudoscalar and one vector meson (see Ref.~\cite{Bhattacharya:2008ke}),
together with corresponding measured branching ratios and $90\%$ C.L. upper bound  
(taken from Ref.~\cite{ParticleDataGroup:2024cfk}) and theoretical predictions
based on best-fit solutions given in Table~\ref{tab:D2PV-best}.
\label{tab:D2PV-CF}}
\centering
\begin{tabular}{c c c c c c}
\toprule
Meson & Mode & Amplitude & $\mathcal{B}_\text{exp}$ ($\%$) & $\mathcal{B}_\text{fit}$  ($\%$) & $\mathcal{B}_\text{fit}$ ($\%$)  \\
& & & & { } & (with widths) \\
 \midrule
$D^0$ & $K^{*-} \pi^+$ & $T_V + E_P$ & $5.34 \pm 0.41$ & $5.25 \pm 0.43$ & $5.28 \pm 0.44$ \\
 & $K^- \rho^+$ & $T_P + E_V$ & $11.2 \pm 0.7$ &   $11.1 \pm 2.0$ &  $11.1 \pm 2.0$ \\
 & $\overline{K}^{*0} \pi^0$ & $\frac{1}{\sqrt{2}}(C_P - E_P)$ & $3.74 \pm 0.27$ & $3.81 \pm 0.42$ &  $3.79 \pm 0.41$ \\
 & $\overline{K}^0 \rho^0$ & $\frac{1}{\sqrt{2}}(C_V - E_V)$ & $1.26 \pm 0.16$ & $1.28 \pm 0.32$ & $1.28 \pm 0.30$ \\
 & $\overline{K}^{*0} \eta$ & $\frac{1}{\sqrt{3}}(C_P + E_P - E_V)$ &  $1.41 \pm 0.12$ & $1.40 \pm 0.20$ &  $1.41 \pm 0.19$ \\
 & $\overline{K}^{*0} \eta\,'$ & $-\frac{1}{\sqrt{6}}(C_P + E_P + 2 E_V)$ & $<0.10$ & $0.009 \pm 0.001$&  $0.177 \pm 0.024$ \\
 & $\overline{K}^0 \omega$ & $-\frac{1}{\sqrt{2}}(C_V + E_V)$ & $2.22 \pm 0.12$ & $2.22 \pm 0.34$ & $2.22 \pm 0.28$\\
 & $\overline{K}^0 \phi$ & $-E_P$ & $0.825 \pm 0.061$ & $0.824 \pm 0.061$ &  $0.824 \pm 0.061$\\
 \midrule
$D^+$ & $\overline{K}^{*0} \pi^+$ & $T_V + C_P$ &  $1.57 \pm 0.13$ & $1.56 \pm 0.25$ &  $1.57 \pm 0.22$ \\
 & $\overline{K}^0 \rho^+$ & $T_P + C_V$ & $12.28 \pm 1.20$ & $14.10 \pm 1.60$& $14.10 \pm 1.50$ \\
 \midrule
$D_s^+$ & $\overline{K}^{*0} K^+$ & $C_P + A_V$ &$3.79 \pm 0.09$ &  $3.79 \pm 0.73$  &  $3.79 \pm 0.74$ \\
 & $\overline{K}^0 K^{*+}$ & $C_V + A_P$ & $1.54 \pm 0.14$ &  $1.65 \pm 0.24$ & $1.65 \pm 0.20$\\
 & $\rho^+ \pi^0$ & $\frac{1}{\sqrt{2}}(A_P - A_V)$ & $-$ &$0.01 \pm 0.05$ &   $0.01 \pm 0.05$\\
 & $\rho^+ \eta$ & $\frac{1}{\sqrt{3}}(T_P - A_P - A_V)$ & $8.9 \pm 0.8$ & $6.7 \pm 0.5$&   $6.8 \pm 0.4$\\
 & $\rho^+ \eta\,'$ & $\frac{1}{\sqrt{6}}(2T_P + A_P + A_V)$ & $5.8 \pm 1.5$ &  $2.4 \pm 0.2$  & $3.5 \pm 0.3$ \\
 & $\pi^+ \rho^0$ & $\frac{1}{\sqrt{2}}(A_V - A_P)$ &  $0.0112 \pm 0.0013$ &  $0.011 \pm 0.047$ &  $0.011 \pm 0.047$ \\
 & $\pi^+ \omega$ & $\frac{1}{\sqrt{2}}(A_V + A_P)$ & $0.238 \pm 0.015$ & $0.239 \pm 0.046$ & $0.239 \pm 0.046$  \\
 & $\pi^+ \phi$ & $T_V$ & $4.50 \pm 0.12$ &  $4.51 \pm 0.12$ & $4.51 \pm 0.12$\\
\bottomrule
\end{tabular}
\end{table}

The CF $D\to PV$ decays and their amplitude decomposition in terms of relevant topological diagrams are given in Table~\ref{tab:D2PV-CF}. Compared to $D\to P_1P_2$ decays, there are more free parameters ($15$ in total) to determine. The notation of the topological amplitude is such that the subscript ($V$ and $P$) on the amplitude specify the final state containing the spectator quark of the decaying charm meson. 

\begin{table}[t]
\centering
\caption{Fit results for $D\to PV $ decays (the amplitudes are
in units of $10^{-6}$ GeV).
\label{tab:D2PV-best}}
\begin{tabular}{c c c}
\toprule
\multirow{2}{*}{\textbf{Parameters}} & \multicolumn{2}{c}{\textbf{Solutions}}  \\
 & no width-effects & with width-effects  \\
\midrule
$T_V$ & $4.00 \pm 0.05$ & $3.99 \pm 0.05$  \\
$C_P$ &$4.05 \pm 0.28$ & $4.00 \pm 0.28$ \\
$\phi_{C_P}$ &  $(-157.70 \pm 1.40) ^\circ$ & $(-157.84 \pm 1.31) ^\circ$ \\
$E_P$ &  $2.86 \pm 0.11$ & $2.85 \pm 0.11$ \\
$\phi_{E_P}$ & $(98.92 \pm 4.14) ^\circ$  &$(99.97 \pm 4.05) ^\circ$ \\
$T_P$ & $7.23 \pm 0.24$ & $6.77 \pm 0.21$ \\
$\phi_{T_P}$ & $(146.43 \pm 7.79) ^\circ$ & $(140.93 \pm 7.48) ^\circ$ \\
$C_V$ & $2.19 \pm 0.14$ & $2.09 \pm 0.10$ \\
$\phi_{C_V}$ &  $(-35.47 \pm 8.46) ^\circ$  &  $(-43.91 \pm 7.51) ^\circ$ \\
$E_V$ &$3.38 \pm 0.16$ &  $3.30 \pm 0.14$ \\
$\phi_{E_V}$ & $(-107.86 \pm 7.36) ^\circ$& $(-112.26 \pm 6.44) ^\circ$ \\
$A_P$ &  $0.41 \pm 0.02$ & $0.41 \pm 0.02$ \\
$\phi_{A_P}$ & $(-52.28 \pm 45.91) ^\circ$  & $(-42.13 \pm 45.13) ^\circ$ \\
$A_V$ &$0.64 \pm 0.02$ & $0.63 \pm 0.02$ \\
$\phi_{A_V}$ & $(-50.02 \pm 24.61) ^\circ$& $(-44.53 \pm 24.80) ^\circ$ \\
\midrule
$\chi^2_\text{min}$ &  $15.87$ & $11.93$  \\
\bottomrule
\end{tabular}
\end{table}

The final state width effects are expected to be pronounced in $D\to PV$ case, as
widths of several vector mesons are not small.
Out of the 18 CF $D$-decays listed in the Table~\ref{tab:D2PV-CF}, branching ratios of 16 modes
is already measured, which allow one to determine $D\to PV$ topological amplitudes
$T_m, C_m, E_m, A_m$ ($m= P, V$) with their strong phases (relative to the amplitude $T_V$)
$\phi_{T_P}, \phi_{C_m}, \phi_{E_m}, \phi_{A_m}$ by performing minimization of $\chi^2$ constructed
similar Eq.~\eqref{eq:chi2}.

In order to see the impact of final state width effects on the parameter extraction,
we present results for both cases: with and without including the width effects in the phase space of decay modes.
We first perform $\chi^2$ test without including the width effects. We find several solutions which
correspond to various local minima of the $\chi^2$ function. In order to avoid local minima solutions,
we perform minimization 500 times with initial guesses for parameters taken from a sample of uniform
random distribution; the amplitudes are taken from  $U(0, 15)$ and phases from $U(-\pi, \pi)$.
The fit results with the least $\chi^2$-minimum value ($\chi^2_\text{min}=15.88$) are listed in the second column
of Table~\ref{tab:D2PV-best}.

Next, we use the values of amplitudes and phases obtained in the fit as initial guesses for parameters to
perform another $\chi^2$-test that also accounts for finite width effects. The solutions obtained after minimizing
the $\chi^2$ function are listed in the final column of Table~\ref{tab:D2PV-best}.
Our analysis reveals that the majority of amplitudes and their phases  do not change significantly.
However, there is an overall reduction in the $\chi^2$-minimum value, decreasing by approximately
$4$ units compared to the case where finite width effects were not considered.
This suggests that incorporating contributions from finite width effects has resulted in an overall improvement
in the fit. 

The most significant consequence of considering the finite width of final-state particles is that the predictions for the branching ratios of charm decay modes that contain $\eta^\prime$ in the final state are notably different now. In Table~\ref{tab:D2PV-CF}, we present the predicted theoretical branching ratios for decays, considering both cases: with and without considering width effects, using solutions presented in Table~\ref{tab:D2PV-best} as input. We note that the in the ``no width-effects" case, the branching ratio of $D^0\to \overline{K}^{\ast 0}\eta^\prime$ decay is expected to be very small, at $0.009\%$. However, when effect of $\overline{K}^{\ast 0}$ width is taken into account, we find  the branching ratio to be $(0.177\pm 0.024)\%$, a strikingly $\sim 20$ times larger value! In case of the decay $D_s^+ \to \rho^+\eta^\prime$, we find theoretical branching ratio to be $(3.5 \pm 0.3)\%$, which is about $46\%$ higher compared to the value obtained without including effects of $\rho^+$ width in the phase space calculation.

On the other hand, predictions for the  branching ratios for the rest of the decay modes remain almost unchanged. This can be understood from the fact that final state particles in both decay modes $D^0\to \overline{K}^{\ast 0}\eta^\prime$ and $D_s^+ \to \rho^+\eta^\prime$ are heavy, leaving very small available phase space for the decay. Inclusion of vector meson widths, which are not small in these cases, leads to more available phase space which is substantial  relatively. But in case of other decay modes, pseudoscalar particles in the final state are either pions or kaons, which are not heavy particles. Therefore, available phase space in these decays is very large  compared to the involved vector meson's width, which explains why width effects are not important in these decays. From this reasoning one can already insinuate that, as we alluded to in Introduction also, such finite width effects will be very important in $D\to VV$ decays, as now both final state particles may have large width as well as mass. We therefore plan to analyze $D\to VV$ decays and assess the impact of final state width effects in a subsequent publication \cite{BKMP:in-progress}.

\section{Conclusions}\label{Conclusions}

The amplitude analyses of two-body nonleptonic charm decays $D\to PP$ and $D\to PV$ are usually performed assuming that final state particles are stable, i.e., ignoring the effects of their finite widths. This might not be the best approximation if the size of the available final state phase space is comparable to the widths of the final state particles.

In this paper, we suggested a method to account for the effects of the non-zero widths of the final state mesons. We applied this method to the CF decays $D\to PP$ and $D\to PV$ to compare the numerical values of the extracted flavor-flow amplitudes in the zero-width approximation to the actual situation of non-zero widths. The resulting decay amplitudes do not significantly differ from the ones obtained in the zero-width approximation, yet, using the obtained amplitudes to predict the branching ratios for the CF $D$-decays, we observe the most substantial difference in the predictions for the $D^0\to \overline{K}^{\ast 0}\eta^\prime$ and $D_s^+ \to \rho^+\eta^\prime$ decays, which is expected for the decays to the final states with restricted phase space. We expect even more substantial effects for $D\to VV$ transitions that will be described in a future publication. We hasten to comment, however, that the finite widths of the final state particles can be treated appropriately if complete Dalitz plot analyses of $D$-decays are performed instead of the studies of two-body decays.

\section{Acknowledgments}
The authors would like to thank John Donoghue for valuable comments and Hai-Yang Cheng for many useful communications. This research was supported in part by the US Department of Energy grant DE-SC0024357.

\appendix
\section{Appendix. Unitarity and unstable particles in the final state}
\label{Appendix}

Many of the $D$-decay processes we discussed have unstable particles in the final state (e.g., $D^0\to K^- \rho^+ $, with $\rho^+$ decaying to $\pi^+\pi^0$). To include the effect of the finite width of these unstable particles, we employ the optical theorem, which is an outcome of the unitarity of the $S$-matrix, to relate the decay rate to the imaginary part of the loop of $1\to 1$ process (see Eq.~\eqref{eq:optical-th}) representing the forward scattering amplitude ${\cal M}(D \to D)$. We then calculate the imaginary part of the processes by cutting propagators of both loop particles (see Fig.~\ref{Fig1}) while working in the narrow-width approximation, which is justified by experimental data. 

However, since the $S$-matrix construction concerns asymptotic states, which are comprised of stable particles only, in principle, one should only cut through stable particles. This means that, for example, in the case $D^0\to K^- \rho^+ $ ($\rho^+\to \pi^+\pi^0$), the cut should be made through $K^-$,  $\pi^+$ and $\pi^0$ in a two-loop diagram similar to Fig.~\ref{fig:2loop}. Below, we give a proof-of-principle example to show the equivalence of both approaches.

Consider a decay $P(p) \to a (p_a) \,b (p_b)$, where $b$ is an unstable particle with width $\Gamma_b$. For simplicity, we assume that it has only one decay mode: $b(p_b) \to c(p_c)\, d (p_d)$. In the following subsections, we first calculate the discontinuity in the dressed propagator of an unstable particle and then show that the discontinuity of one loop diagram with cut-through particles $a$ and $b$ is the same as that of two loop diagrams with cut-through particles $a$, $c$, and $d$. For simplicity, we will assume that all particles involved are scalars. 

\subsection{Discontinuity in the propagator}
\begin{figure}[t]
\centering
     \includegraphics[width=0.8\textwidth]{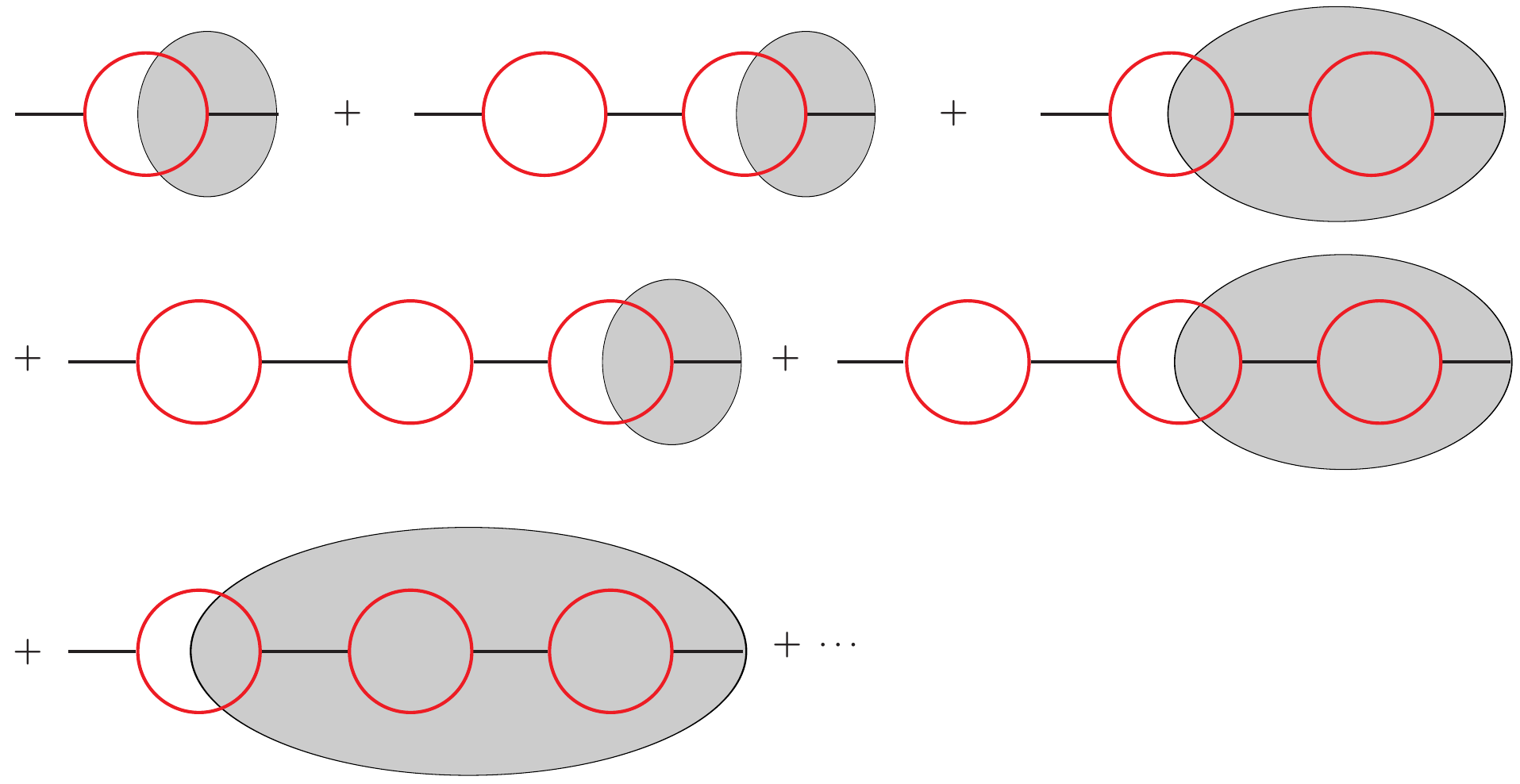}
\caption{Diagrams showing cuts in the propagator.}
\label{fig:bubble_sum}
\end{figure}

The discontinuity in the dressed propagator of unstable scalar particle is given by the sum of the diagrams shown  in Fig. \ref{fig:bubble_sum}, where the shaded region indicates the cut in the diagram\footnote{For a more detailed exposition of the proof, we refer to Ref.~\cite{Donoghue:2019fcb}.} with the quantities in the shaded region being the Hermitian conjugates of the quantities in the unshaded region. As there is no perturbative expansion for the decay amplitudes, all diagrams must be summed over.

We denote free scalar propagator (shown as black lines) as
\begin{align}
	i S_0 (q) = \frac{i}{q^2 - m_0^2 + i \epsilon}\,,
\end{align}
and the amplitudes of basic one-loop diagram without cut and with cut as $-i \Sigma(q)$ and $-i\operatorname{Disc}\Sigma(q)$, respectively. Note that all quantities in the shaded region are to be complex-conjugated. The sum of all the diagrams $iS(q)$ can be written as (suppressing the momentum dependence)
\begin{eqnarray}\label{eq:disc_dressed}
	 \operatorname{Disc} i S(q) &=& 
     iS_0 (-i\operatorname{Disc}\Sigma) (iS_0)^\ast
     \nonumber \\
	 &+& iS_0(-i\Sigma) iS_0 (-i\operatorname{Disc}\Sigma) (iS_0)^\ast  + iS_0(-i\operatorname{Disc}\Sigma)(iS_0)^\ast (-i\Sigma)^\ast (iS_0)^\ast\nonumber\\
	 &+& iS_0 (-i\Sigma) iS_0 (-i\Sigma) iS_0(-i\operatorname{Disc}\Sigma)(iS_0)^\ast 
	 + iS_0 (-i\Sigma)i S_0(-i\operatorname{Disc}\Sigma)(iS_0)^\ast (-i\Sigma)^\ast (iS_0)^\ast\nonumber\\
	  &+& iS_0(-i\operatorname{Disc}\Sigma)(iS_0)^\ast (-i\Sigma)^\ast(iS_0)^\ast (-i\Sigma)^\ast(iS_0)^\ast + \cdots
      \\
	 &=& i S_0\left(1 + \Sigma S_0 + (\Sigma S_0)^2 + \cdots \right)(-i\operatorname{Disc}\Sigma) \left(1 + S_0^\ast \,\Sigma^\ast + (S_0^\ast \,\Sigma^\ast)^2 + \cdots \right)(iS_0)^\ast 
     \nonumber\\
	 &=& \frac{i S_0}{1- \Sigma\, S_0 }(-i\operatorname{Disc}\Sigma)\frac{(iS_0)^\ast}{1 - \Sigma^\ast\, S_0^\ast}
	 = \left(\frac{i}{q^2 - m_0^2 - \Sigma}\right)(-i\operatorname{Disc}\Sigma)\left(\frac{i}{q^2 - m_0^2 - \Sigma^\ast }\right)^\ast, \nonumber
\end{eqnarray}
where now the propagators on each side are the dressed scalar propagator. The one loop amplitude $\Sigma$ contain both real and imaginary parts. The real part of  $\Sigma$ shift the bare mass $m_0$ of the propagator; the physical mass ($m$) is then defined by the solution of the equation $\left(q^2 - m_0^2 - \operatorname{Re}\Sigma\right)_{q^2=m^2} = 0$. On the other hand, the imaginary part is related to the width of the particle, as shown below.

Calculating the one loop amplitude $\Sigma$ in the dimensional regularization scheme in dimension $d= 4- \epsilon$ dimension, we obtain 
%
\begin{align}
	-i\Sigma = \frac{ig^2}{16 \pi^2}\int_0^1 dx \left(\frac{2}{\epsilon} - \gamma_E - \log\frac{m_1^{2} -x (m_1^2-m_2^2 + q^2(1-x)) -i \epsilon}{4 \pi \mu^2}\right),
\end{align}
where $g$ denotes the vertex coupling and $m_{1, 2}$ the mass of the loop particles, and $\gamma_E$ is the Euler-Mascheroni constant. Using the identity $\log (-A) \equiv \log |A| - i \pi $, we obtain
\begin{align}
	\Sigma &= -\frac{g^2}{16 \pi^2}\Bigg[ \int_0^1 dx \left(\frac{2}{\epsilon} - \gamma_E - \log\left|\frac{x (m_1^2-m_2^2 + q^2(1-x)) -m_1^{2}}{4 \pi \mu^2}\right|\right)\nonumber\\
	& \hskip 2cm +~ i \pi \int_0^1 dx ~\theta\{x (m_1^2-m_2^2 + q^2(1-x)) -m_1^{2}\}\Bigg]\nonumber\\
	&= -\frac{g^2}{16 \pi^2}\Bigg[\int_0^1 dx \left(\frac{2}{\epsilon} - \gamma_E - \log\left|\frac{x (m_1^2-m_2^2 + q^2(1-x)) -m_1^{2}}{4 \pi \mu^2}\right|\right) \nonumber\\
	& \hskip 2cm +~ i \pi  \frac{\sqrt{\lambda(m^2, m_1^2, m_2^2)}}{m^2} \theta (m - m_1-m_2)\Bigg]\nonumber\\
	& = -\frac{g^2}{16 \pi^2}\int_0^1 dx \left(\frac{2}{\epsilon} - \gamma_E - \log\left|\frac{x (m_1^2-m_2^2 + q^2(1-x)) -m_1^{2}}{4 \pi \mu^2}\right|\right)
		- i \,m\, \Gamma\,,
\end{align}
%
evaluated at $q^2 = m^2$. In the last term, $\Gamma =\sqrt{\lambda(m^2, m_1^2, m_2^2)}/16\pi m^3 $ denotes the decay width  of initial particle.

From the discussion delineated above, and using identity $\operatorname{Disc} A= 2 i \operatorname{Im} A$, we can rewrite Eq.~\ref{eq:disc_dressed} in a simplified form as
\begin{align}\label{eq:disc_fullpropagator}
	&\left(\frac{i}{q^2 - m^2  + i m \,\Gamma_\text{total}}\right) ( 2 \operatorname{Im}  \Sigma) \left(\frac{i}{q^2 - m^2  + i m\, \Gamma}\right)^\ast\nonumber\\
	&\quad = iS(q) \, 2 m\Gamma \, (iS(q))^\ast.
\end{align}
The equation gives the discontinuity in the full scalar propagator $S(q)$.

It would be assuring to verify that setting  $\Gamma\to 0$ in Eq.~\ref{eq:disc_fullpropagator}, one should obtain
 cutting rule of stable particle particle propagator. Eq.~\ref{eq:disc_fullpropagator} can be written as
 \begin{align}
 	 iS(q) \, 2 m\Gamma \, (iS(q))^\ast  = \frac{2 m \Gamma}{(p^2-m^2)^2 + (m\Gamma)^2}.
 \end{align}
Now, taking limit $\Gamma\to 0$, and using the identity:
\begin{align}
	\lim_{\alpha \to 0} \frac{\alpha}{x^2 + \alpha^2} = \pi \delta(x),
\end{align}
we obtain
 \begin{align}
 	 \lim_{\Gamma \to 0} \left[iS(q) \, 2 m\Gamma \, (iS(q))^\ast\right]  = 2\pi \delta(p^2 -m^2),
 \end{align}
which is the correct cutting rule for the propagator $i/(p^2 -m^2)$. 

\subsection{Cutting through unstable particles}
\begin{figure}[h]
\centering
     \includegraphics[width=8cm]{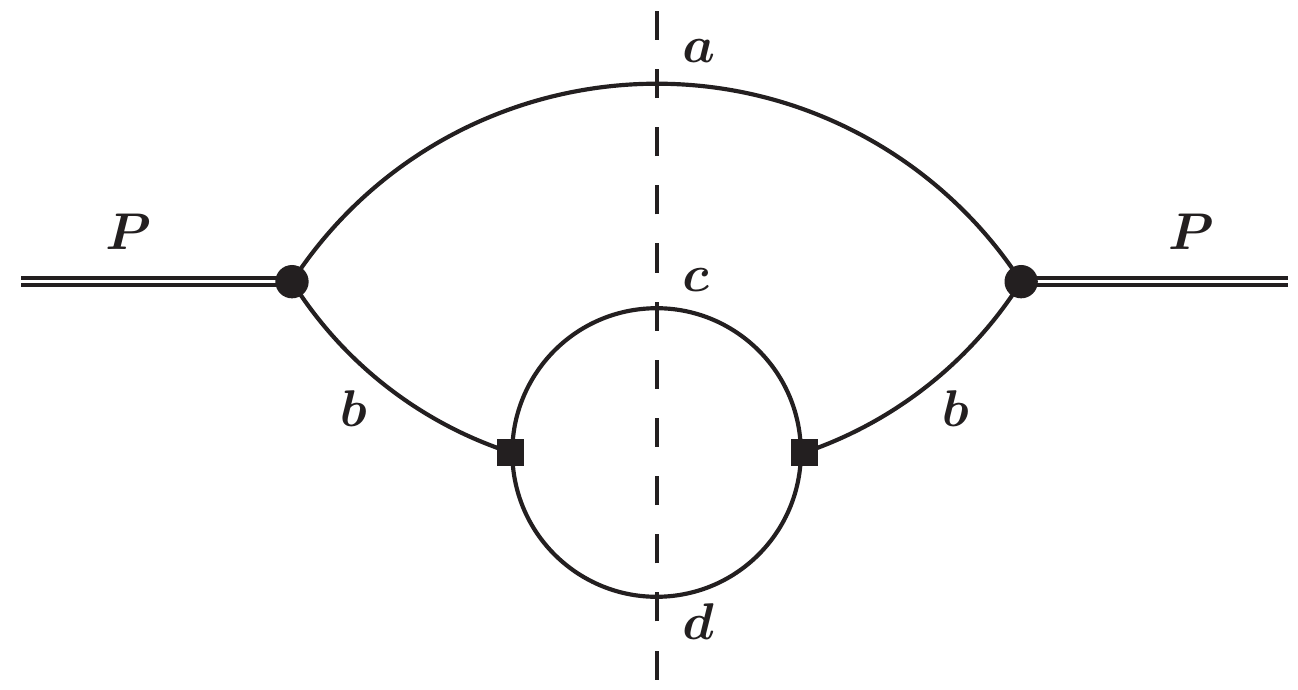}
\caption{Imaginary part of the amplitude squared for the decay $P\to a\, b\, (\to c\,d)$.}
\label{fig:2loop}
\end{figure}
%

The discontinuity in the amplitude is
\begin{eqnarray} 
	2 \operatorname{Im} \mathcal{M}  &=& \int \frac{d^4 p_b}{(2\pi)^4}\int \frac{d^4 p_d}{(2\pi)^4} \, i \lambda \,i D(p_b)  \,i \kappa \, 2\pi \delta (p_d^2 - m_d^2) \theta((p_d)_0)\,
	 2\pi \delta ((p_b - p_d)^2 - m_c^2) \nonumber\\ 
    && \theta((p_b - p_d)_0)\ 
	(i \kappa)^\ast \left(i D(p_b)\right)^\ast \, (i \lambda)^\ast \, 2\pi \delta ((p - p_b)^2 - m_a^2) \theta((p - p_b)_0)\,,
\end{eqnarray}
where $m_x$ denotes mass of the $x$ particle, $\lambda$  and $\kappa$ are the decay amplitudes associated with $P \to ab$ and $b \to cd$ decays, respectively and can be treated as ``vertices" in the corresponding diagrams. The total width of the particle $b$ is given by the decay $b \to cd$. Note that, experimentally, this decay is not detected when a two-body transition $P \to ab$ is studied, but it defines the width $\Gamma_b$.
Here $i\,D(p_b)$ denotes the propagator for $b$ particle,
\begin{align}\label{eq:b_propagator}
	i\, D(p_b) &= \frac{i}{p_b^2 - m_b^2 + i m_b \Gamma_b}\,,
\end{align}
where $\Gamma_b$ is the width of $b$ particle.\\

For further simplification, we insert a factor of $1 = \int d^4 p_c \,\delta^4(p_b-p_d-p_c)$ to arrive at
\begin{align} 
	2 \operatorname{Im} \mathcal{M}  
	 &= \int \frac{d^4 p_b}{(2\pi)^4}\,  \lambda \, iD(p_b) \int \frac{d^4 p_d}{(2\pi)^4} \int \frac{d^4 p_c}{(2\pi)^4} (2\pi)^4 \delta^4(p_b-p_d-p_c) \,|\kappa|^2 \nonumber\\
	 &\hskip2cm\times  ~~  2\pi \delta (p_d^2 - m_d^2) \theta((p_d)_0)\,
	\, 2\pi \delta ((p_b - p_d)^2 - m_c^2) \theta((p_b - p_d)_0)\nonumber\\
	&\hskip2cm\times  ~~  \left(i D(p_b)\right)^\ast \,  \lambda^\ast \, 2\pi \delta ((p - p_b)^2 - m_a^2) \theta((p - p_b)_0)\,.
\end{align}

Integrating over the delta functions using the identity:
\begin{align}
	\int \frac{d^4 q}{(2\pi)^4} 2\pi \delta(q^2 - m^2 ) \theta(q_0) = \int \frac{d^3 q}{(2\pi)^3} \frac{1}{2 \omega_q},
\end{align}
and identifying the $b\to c d$ matrix element as $\mathcal{M}_{b\to cd} = \kappa $, we obtain
\begin{align}
	2 \operatorname{Im} \mathcal{M}  &= \int \frac{d^4 p_b}{(2\pi)^4}\, \lambda \, iD(p_b)\left\{\int \frac{d^3 p_d}{(2\pi)^4\, 2\omega_{p_d}} \int \frac{d^4 p_c}{(2\pi)^4\,2\omega_{p_c}} (2\pi)^4 \delta^4(p_b-p_d-p_c) |\mathcal{M}_{b\to cd}|^2\right\} \nonumber\\
	&\hskip4cm\times  ~~  \left(i D(p_b)\right)^\ast \,  \lambda^\ast \, 2\pi \delta ((p - p_b)^2 - m_a^2) \theta((p - p_b)_0)\,,
\end{align}
The factor in the curly brackets  in the first line is basically the expression of the $2m_b$ times the decay width of $b\to cd $. For simplicity, let's assume that $b$ decays to $cd$ only.
Then, the above expression becomes
\begin{align}
	2 \operatorname{Im} \mathcal{M}  &= \int \frac{d^4 p_b}{(2\pi)^4}\, \lambda
	\left[iD(p_b)2m_b \Gamma_{b} \left(i D(p_b)\right)^\ast\right]  \lambda^\ast \, 2\pi \delta ((p - p_b)^2 - m_a^2) \theta((p - p_b)_0)\,,
\end{align}
%
From the  discussion in the previous subsection we recognize that
$\left[iD(p_b2m_b \left(i D(p_b)\right)^\ast\right]$
describes the discontinuity of propagator $i D(p_b)$:
\begin{align}
	\operatorname{Disc}iD(p_b) = i D(p_b)  2m_b\Gamma_{b}\left(i D(p_b)\right)^\ast,
\end{align}
%
Using the above relation, we finally obtain
\begin{align}
	2 \operatorname{Im} \mathcal{M} &= \int \frac{d^4 p_b}{(2\pi)^4}\, 
	 |\lambda|^2 \,\operatorname{Disc}iD(p_b)\,2\pi \delta ((p - p_b)^2 - m_a^2) \theta((p - p_b)_0)\,,
\end{align}
which is the same expression one would get if one calculates discontinuity in the $P\to a b \to P$ diagram
similar to Fig.~\ref{Fig1}, treating $b$  as stable particle.

\bibliography{refs}{}
\bibliographystyle{utcaps_mod}

\end{document}